\documentclass[a4paper]{article}

\usepackage{icrc2013}

\title{Night Sky Background Analysis for the Cherenkov Telescope Array using the Atmoscope instrument}

\shorttitle{NSB Analysis for the CTA using the Atmoscope}

\authors{
M. Gaug$^{1,2}$
for the CTA Consortium.
}

\afiliations{
$^1$ F{\'i}sica de les Radiacions, Departament de F{\'i}sica, Universitat Aut{\`o}noma de Barcelona, 08193 Bellaterra, Spain. \\
$^2$ CERES, Universitat Aut{\`o}noma de Barcelona-IEEC, 08193 Bellaterra, Spain. \\
}

\email{markus.gaug@uab.cat}

\abstract{The site selection group for the future Cherenkov Telescope Array (CTA)
has deployed sensitive light sensors at 9 candidate sites, 5 of them in the Southern and 4 in the Northern hemisphere. 
The sensors are equipped with a PIN diode and a calibrated V-filter, and a
blue/UV filter matching the spectral response of the photomultipliers to
be employed in the CTA cameras. All sensor installations, denominated
"Atmoscopes", have been cross-calibrated before deployment, and their
sensitivity is monitored every two to five months. We show that a thoroughly
developed model of the integral contribution of starlight to the overall
light measure serves as an additional cross-calibration for each device
during each night, reducing the systematic uncertainty of this measurement
to less than 15\%. The starlight can then be subtracted from the
measurements, and the residuals compared among the different sites. We
show that in most cases a decomposition into the contributions from
zodiacal light, airglow and anthropogenic light pollution is possible.
}


\begin{document}
\maketitle

\section{Introduction}

Measuring the light-of-night-sky  is a difficult task at dark astronomical sites. The small amount of residual light requires 
sensitive and well-calibrated instruments with a large duty-cycle, in order to assess correctly hourly up to seasonal variations. 
Although sensitive optical telescopes, equipped with a set of standard filters, observing a reference star and its surroundings~\cite{roach,walker,leinert,patat},
 are best suited for that task, these are relatively expensive, and existing telescopes are typically not available for an extended 
site characterization campaign. 

The CTA~\cite{ctaconcept,bulik} site search requires to assess the quality of the night sky at 9 candidate sites, covering typically one year of data 
for each, at wavelengths typical for the Cherenkov light emitted by $\gamma$-ray induced air showers, to which the photo-detection devices in the 
cameras as well as  all optical components (mirrors, protecting plexiglas and light concentrators), are matched. Hence the backgrounds of the UV 
to visible range 
of the light spectrum, from about 300~nm to 600~nm wavelength need to be characterized. 

Commercially available devices include \textit{sky-quality-meters} (SQM)~\cite{sqm,cinzano,puerto}, 
\textit{all-sky-cameras}~\cite{duriscoe,cinzanofalchi} and \textit{astronomic monitors} (AstMon)~\cite{puerto}. 
The first two have the disadvantage to accept light over a broad energy band, 
extending from blue to infrared, and therefore only suited for rough cross-calibrations. 
Especially acceptance in the red up to $>$700~nm, and infrared leakage, can distort 
the information about the true light-of-night-sky leaking into the 
photo-detectors of the CTA cameras~\cite{ctacameras}. 
Another disadvantage of the SQM is its large field-of-views with considerable acceptance of light until incidence angles of almost 80$^\circ$. 
Since the light-of-night-sky typically increases towards the horizon, but Cherenkov telescopes do not observe at such low altitudes, 
results may be distorted by horizon effects. Finally, measurement campaigns with adapted spectrometers have been carried out~\cite{aube}.

The members of CTA decided to construct a new apparatus, equipped with a sensitive PIN diode, a filter pair matching the 
blue/UV response of the photomultipliers, and a standard V-filter, and a lens limiting the field-of-view to roughly 50$^\circ$ in diameter.
The advantage of this approach is that 10 instruments could be equipped with exactly the same hardware, 
the boxes can be dismounted at any time and the whole or individual components cross-calibrated frequently. This yields at least a relative precision 
of the measurements between the investigated candidate sites unachievable by other means. 

The light-of-night-sky sensor is always pointing to zenith with a relatively large field-of-view. This means that stars are always shining into 
the sensor, their contribution can make up more than half of the overall light yield. Thus, in order to compare the quality of individual 
sites in a fair way, a method has been developed to subtract the effect of the star light, the light from planets, and of zodiacal light. The remaining part can then 
be investigated for inner-night and seasonal variations. 

This proceeding describes how adequate starlight and zodiacal light models are produced for each site and the achieved precision with that method.

\begin{figure}[h!t]
\hspace{-1.2cm}
\centering
\includegraphics[width=0.3\textwidth]{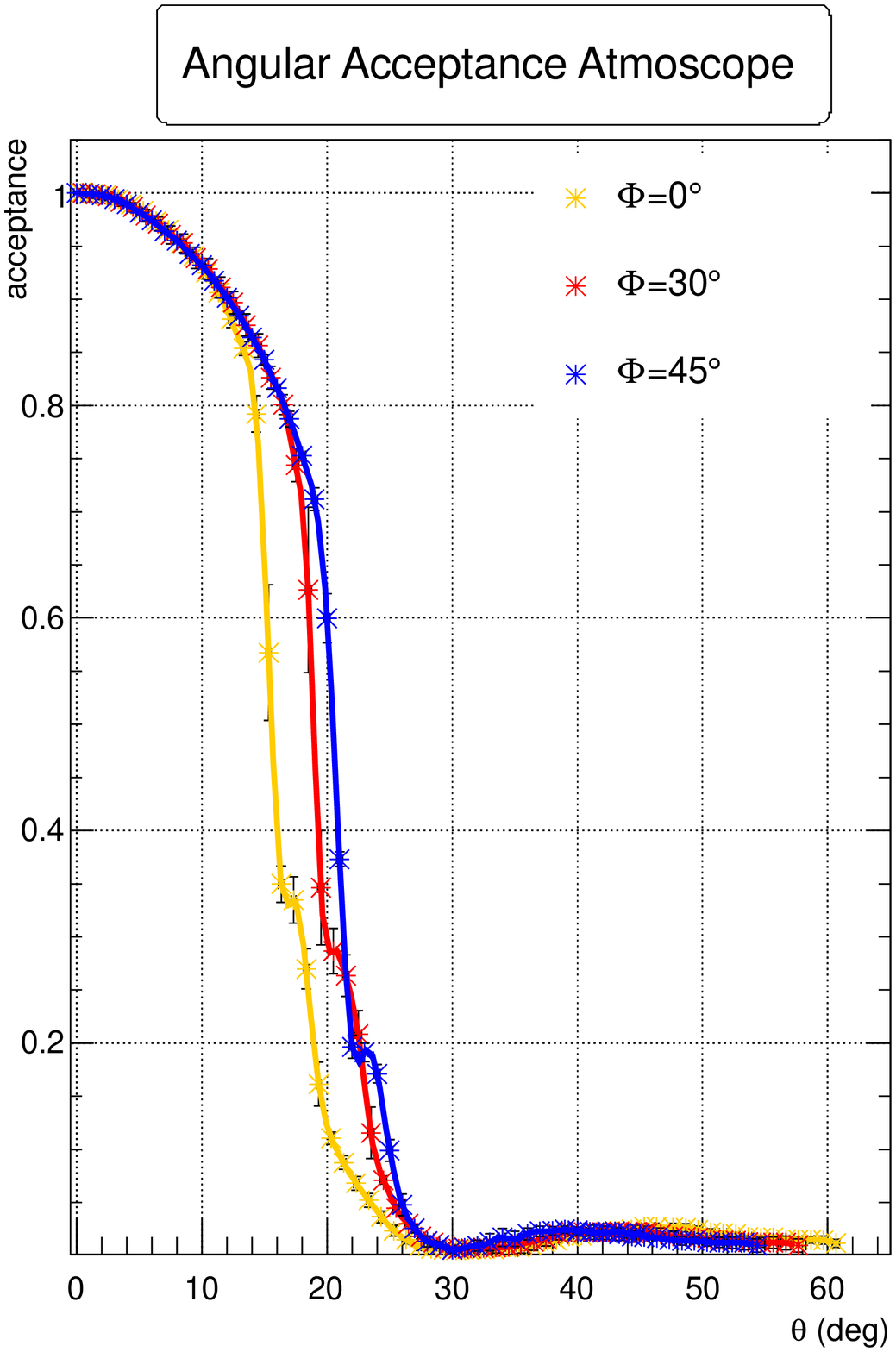}
\includegraphics[width=0.38\textwidth]{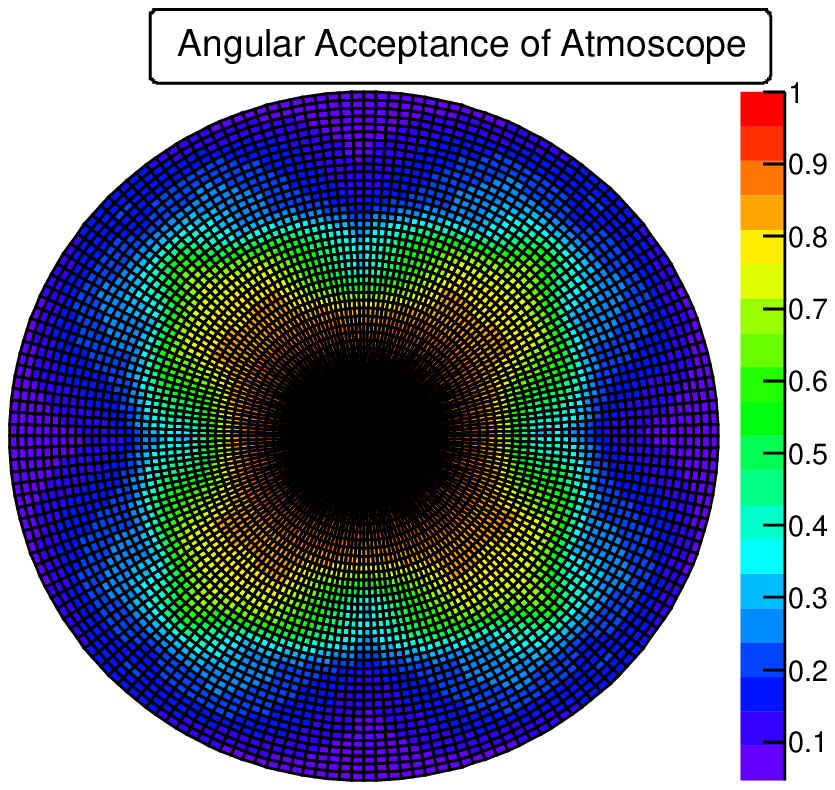}
\vspace{-0.6cm}
\caption{\small{Top: angular acceptance of the atmoscope, measured parallel to the PIN diode ($\Phi=0^\circ$), along its diagonal ($\Phi=45^\circ$)
 and at an intermediate angle ($\Phi=30^\circ$), bottom: in polar coordinates up to an incidence angle of 30 degrees. \label{fig:acceptance}
\vspace{-0.6cm}}}
\end{figure}

\vspace{-0.3cm}
\section{The Atmoscope Light-of-Night-Sky sensor}

The Atmoscope is equipped with a sensitive light sensor, consisting of a lens to limit the field-of-view, 
two different colour filters matching the visible and the blue-UV range, and a $28 \times 28$ mm Silicon PIN diode, 
operated in photovoltaic mode. 
Due to the quadratic shape of the PIN diode, the acceptance of the light sensor is not symmetric in azimuth, as 
shown in fig.~\ref{fig:acceptance}. The device has a total field-of-view of 0.367$^{+0.002}_{-0.000}$~Sr. 
The very small uncertainty is statistical only, assuming perfect 8-fold symmetry of the system and doubling of the measurement uncertainty at every 10$^\circ$
distance to the closest azimuth measurement. 
The measurements have been made on one atmoscope in the lab, 
and degradations of the acceptance due to dirt, aging, small imperfections of the symmetry, etc. are not included here. 

\begin{figure}[h!t]
\centering
\includegraphics[width=0.45\textwidth]{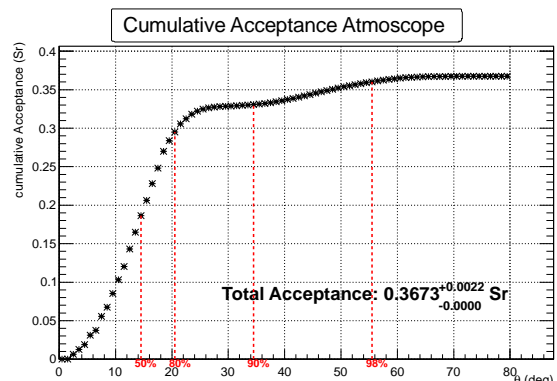}
\vspace{-0.5cm}
\caption{\small{Cumulative angular acceptance of the atmoscope. The red lines show the incidence angles up to which 50\%, 80\%, 90\% and 98\% of the 
total acceptance has been integrated. \label{fig:acceptancecum}
\vspace{-0.4cm}}}
\end{figure}

About 20\% of the acceptance is obtained for zenith angles greater than 20$^\circ$, and a residual of 10\% for zenith angles greater than 35~degrees. 
This analysis integrates light up to 55$^\circ$, which is precise to about 2\% (see fig.~\ref{fig:acceptancecum}). 
Several atmoscopes have a shadowing mast in the field-of-view of the light sensor, which may decrease the amount of registered light by 
another 1\%. 

\begin{figure}[h!t]
\centering
\includegraphics[width=0.95\columnwidth]{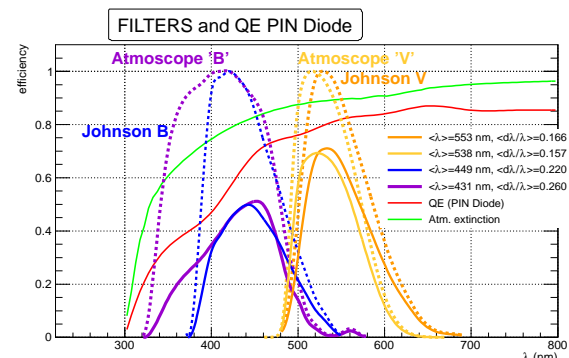}
\vspace{-0.5cm}
\caption{\small{Transmission curves of the PIN diode (red), the atmoscope filters (violet and orange) and the atmospheric transmission for a typical 
clear astronomic night. The blue and yellow lines show the standard Johnson/Bessell B and V-filters for comparison. The dotted lines 
show the normalized transmission of the filters only, the full lines after convolution with the spectral acceptance of the PIN diode and 
the atmospheric transmission. \label{fig:filters}}}
\end{figure}

The filters used in the atmoscope match more or less the standard Johnson/Bessell V and B-filters~\cite{johnson,bessell,bessell1}, however with some 
shift of the mean wavelength. Our chosen blue/UV filter combination shows a broader spectral acceptance than the standard filter (see fig.~\ref{fig:filters}). 
The effect of these shifts need to be corrected in the analysis later on, or in order to compare results with other instruments. 
Peak transmissions range from 0.85 to 0.88 for the visible band and from 0.61 to 0.71 in the blue/UV band. Special care has been taken to control 
filter leakage at other wavelengths, especially in the near infrared, where the atmosphere is much brighter. While the V-filter shows leakage
always below the $10^{-4}$ level, and the contribution of leaking light from longer wavelengths can be conservatively estimated to always below 
0.1\,\% of the light registered inside the nominal window, the blue/UV filter shows stronger leakage. A small window around 550~nm blocks light only to 
a level of $10^{-2}$, and another larger leak lets pass through $10^{-3}$ of the light at wavelengths around 700~nm. We conservatively estimate that 
the first leak may contribute up to 2\,\% of the nominal filter wavelengths. 

\begin{figure}[h!t]
\centering
\includegraphics[width=0.48\textwidth]{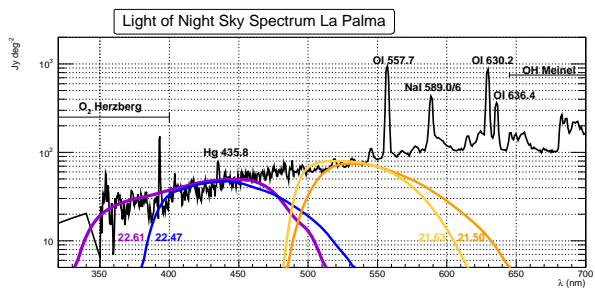}
\vspace{-0.8cm}
\caption{\small{Example of a night-sky spectrum from an astronomical site, here the Observatorio del Roque de los Muchachos at La Palma~\protect\cite{benn}, 
folded with the spectral acceptance of the PIN diode and the two filters used in the atmoscope, and the two standard Johnson/Bessell 
V~and B-filters for comparison. The numbers below the filter curves show the integral in units of magnitude per square arc-second.\label{fig:lons}
\vspace{-0.5cm}}}
\end{figure}

Fig.~\ref{fig:lons} shows how the filters act on a typical light-of-night-sky spectrum at an astronomical site. A shift of 0.14 magnitudes 
is observed for the filter used in the atmoscope, if compared with a standard Johnson/Bessell B-filter, and a shift of 0.13 magnitudes in 
the case of the V-filter, in both cases towards higher magnitudes, i.e. lower levels of background light. As the spectrum of the residual 
light may differ from site to site, these numbers may rather give an order of magnitude for the systematic offset of the atmoscope results, if 
compared with measurements by e.g. a telescope equipped with the standard filters. 

Most atmoscope light sensors have been deployed vertically, i.e. pointing directly to zenith, however some of them show deviations 
from verticality up to 9$^\circ$. Additionally, all light sensors have been installed such that the edges of the  PIN~diode point exactly 
to the geographic North. In a few stations, deviations from this behavior orientation occur, which need to be corrected. Finally, the 
internal clocks of the atmoscopes show considerable drifts, advancing about $2\cdot 10^{-5}$~s/s, with a spread of $5\cdot 10^{-6}$ between 
different clocks. The clocks are synchronized with a GPS reference from time to time, the parts between synchronizations 
are corrected assuming a linear drift with time. 

The sensitivity of the sensors, and especially the lenses, is monitored from time to time with a calibration device emitting continuous light 
at various, controlled intensities. Typically, a calibration is performed directly before and after cleaning the lens, and the difference attributed 
to dust deposits on the lens. Then, a linear interpolation is made in time, between the reduced sensitivity from before cleaning, 
and the one obtained from the previous calibration after cleaning. Typical correction factors range from zero to eight percent for periods of 
several months.


Data have been reduced to exclude Sun and Moon light, humidity above 85\%, and clouds by using information from a commercial thermopile
installed at each atmoscope.
\vspace{-0.6cm}

\section{Contributions of Star and Zodiacal light}

As a next step, the absolute contribution of star light to the measurements was modeled, and compared with data. Astrometry and photometry of the 
\textit{ASCC-2.5} catalog~\cite{kharchenko} was used, which is mainly based on the \textit{Tycho-2 catalog}~\cite{tycho}, with magnitudes converted 
to the Johnson filter system and accurate to better than 0.05$^m$.
Additional light from stars with magnitudes greater than $\sim$12$^m$, 
and the contribution of galaxies have been modeled with the help of the \textit{Nomad catalog}~\cite{nomad}, 
precise to about 0.25$^m$~\cite{monet}. The combination of both catalogs is supposed to be complete to magnitude 20$^m$, 
All catalog entries have been treated as point sources, including the galaxies.

\begin{figure}[h!t]
\centering
\hspace{-0.55cm}
\includegraphics[width=0.35\textwidth]{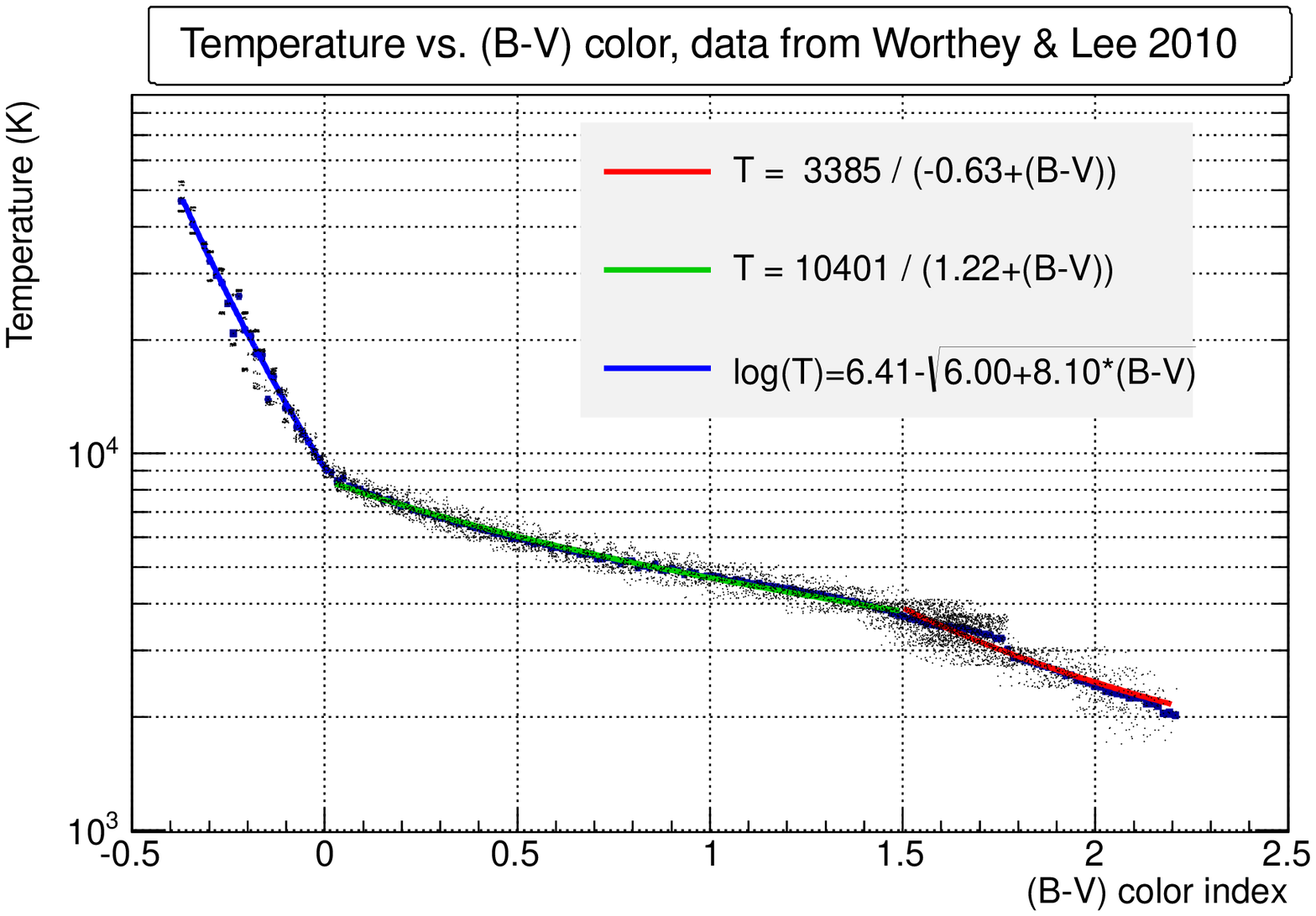}
\includegraphics[width=0.38\textwidth]{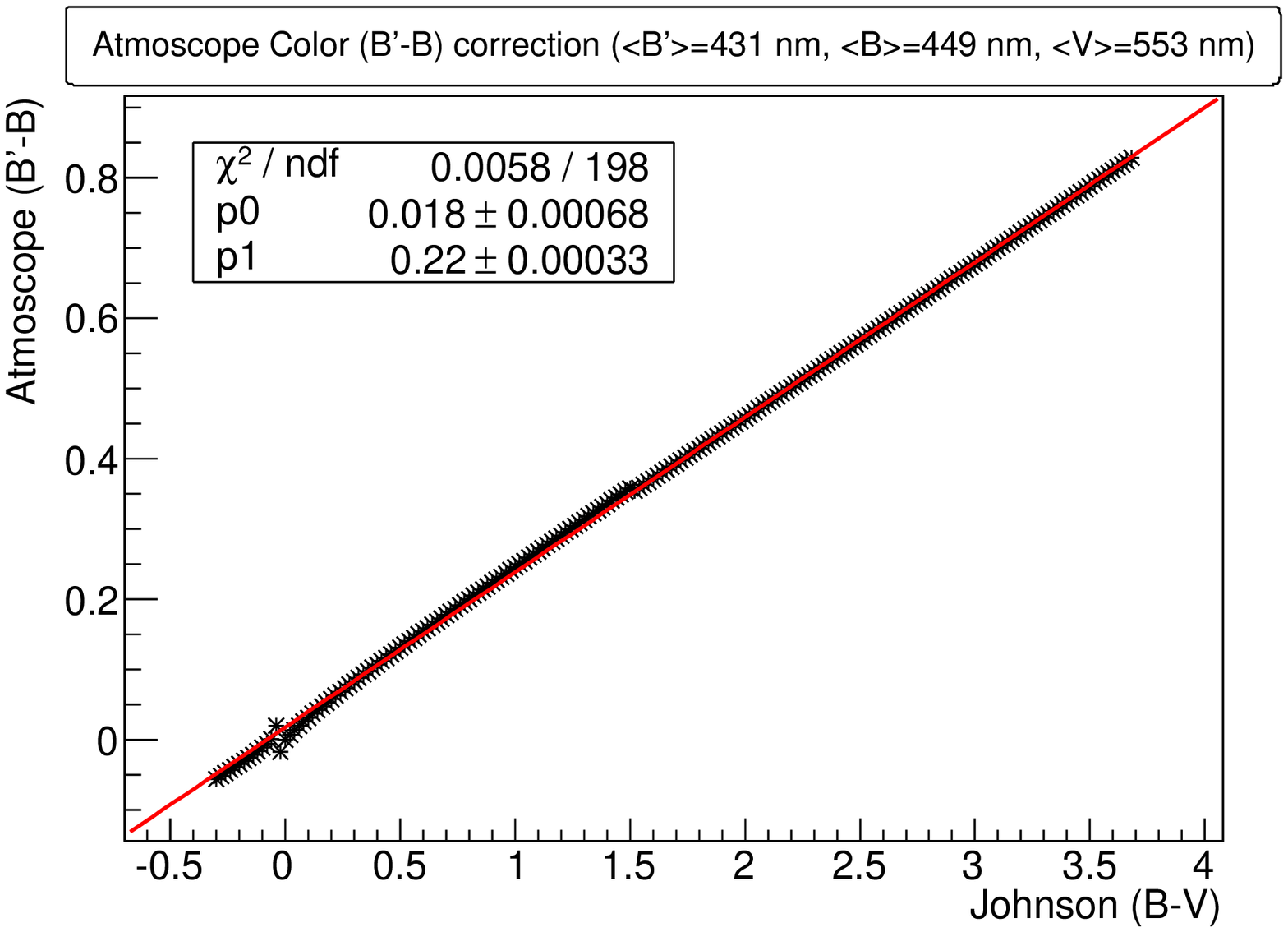}
\vspace{-0.5cm}
\caption{\small{Top: dependency of simulated star temperatures vs. the observed Johnson (B-V) color index. 
The individual points show all simulated star metallicities, their averages being fit to the 3 functions shown in the legend.
Bottom: the modeled Atmoscope (B'-B) color correction w.r.t. the Johnson (B-V) color index. 
Each point corresponds to a different surface temperature of a main sequence star. 
The relation should be, approximately, valid for un-reddened spectra.\label{bminv}
\vspace{-0.4cm}}}
\end{figure}

A dedicated correction had to be applied for the atmoscope B-filter.
Since we did not have a telescope at hand which could observe reference stars with both filters, 
a theoretical prescription was searched to convert a pair of B~and V-filter magnitudes to an atmoscope-B magnitude. 
In a first step, (B-V) color indices were converted to star temperatures, using recent star atmosphere models~\cite{worthley}. The result was fitted 
by adequate functions over 3 separate ranges of the color index (see fig.~\ref{bminv} (top)).

Based on these functions, a (B'-B) correction prescription was obtained, assuming pure black-body emission from the star envelopes, 
and mean filter wavelengths (see fig.~\ref{bminv} bottom):
\begin{eqnarray}
  (B'-B) &=& (0.018\pm0.002) +  (0.221\pm0.015) * (B-V) \nonumber\\
{} && \qquad
.   
\label{eq.bminb}
\end{eqnarray}
In order to convert the magnitudes to photon fluxes, the STIS spectrum of Vega~\cite{bohlin}, and a specially tailored model long-ward of 5337\,\AA~\cite{kurucz}
was used. This model is claimed to be precise to 1\% by the authors.

\begin{figure}[h!t]
\centering
\includegraphics[width=0.45\textwidth]{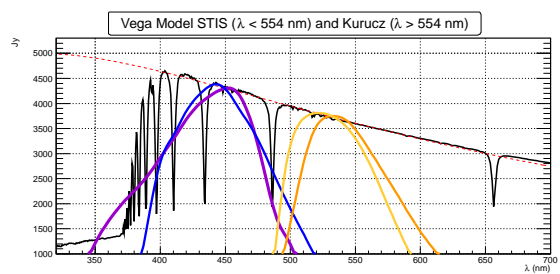}
\vspace{-0.4cm}
\caption{\small{The spectrum of Vega, from~\protect\cite{bohlin,kurucz}. In colors shown are filter acceptances, 
folded with the spectral acceptance of the PIN diode and a typical atmospheric extinction curve. in blue and orange: standard Johnson B and V filters, 
in violet and yellow: the filters used for the atmoscope.\label{vega}
\vspace{-0.2cm}}}
\end{figure}

Fig.~\ref{vega} shows the obtained Vega spectrum together with its coverage by the Johnson B and V-filters, 
and those used in the atmoscope. The spectrum was then integrated following the advice of~\cite{castelli}: 
since the photo-diode acts as a photon-counting device, the integration of the spectrum takes the form:
\begin{equation}
\int f_\nu (\lambda)/\lambda  \cdot R(\lambda) d\lambda~,
\end{equation}
where $f_\nu$ stands for the source brightness and $R(\lambda)$ for the normalized system response function, 
including filter transmission, PIN diode quantum efficiency and atmospheric transmission. 
The integration of the spectrum 
with the Johnson V-filter curve yields results compatible with the values found in the literature 
to better than 1\%, for the Johnson B-filter to better than 2\%. 
If only a spectral point at the mean filter wavelength is used, and multiplied with the normalized integrated filter width, 
compatible results for the V filter are obtained, but slightly higher values for the Johnson-B filter and much higher values for the Atmoscope B-filter. 
The reason for this discrepancy are the atomic absorption lines, mainly around 434~nm and 410~nm. 
The effect of these lines amounts to 7\% for the Johnson-B filter and 24\% for the Atmoscope-B filter. 
 
Atmospheric extinction was applied to the resulting photon fluxes using published median B and V-filter extinction values for all sites. 
A similar procedure has been applied to the contribution of planets, until Neptune. Satellites are not yet taken into account.

\begin{figure}[h!t]
\centering
\includegraphics[width=0.48\textwidth]{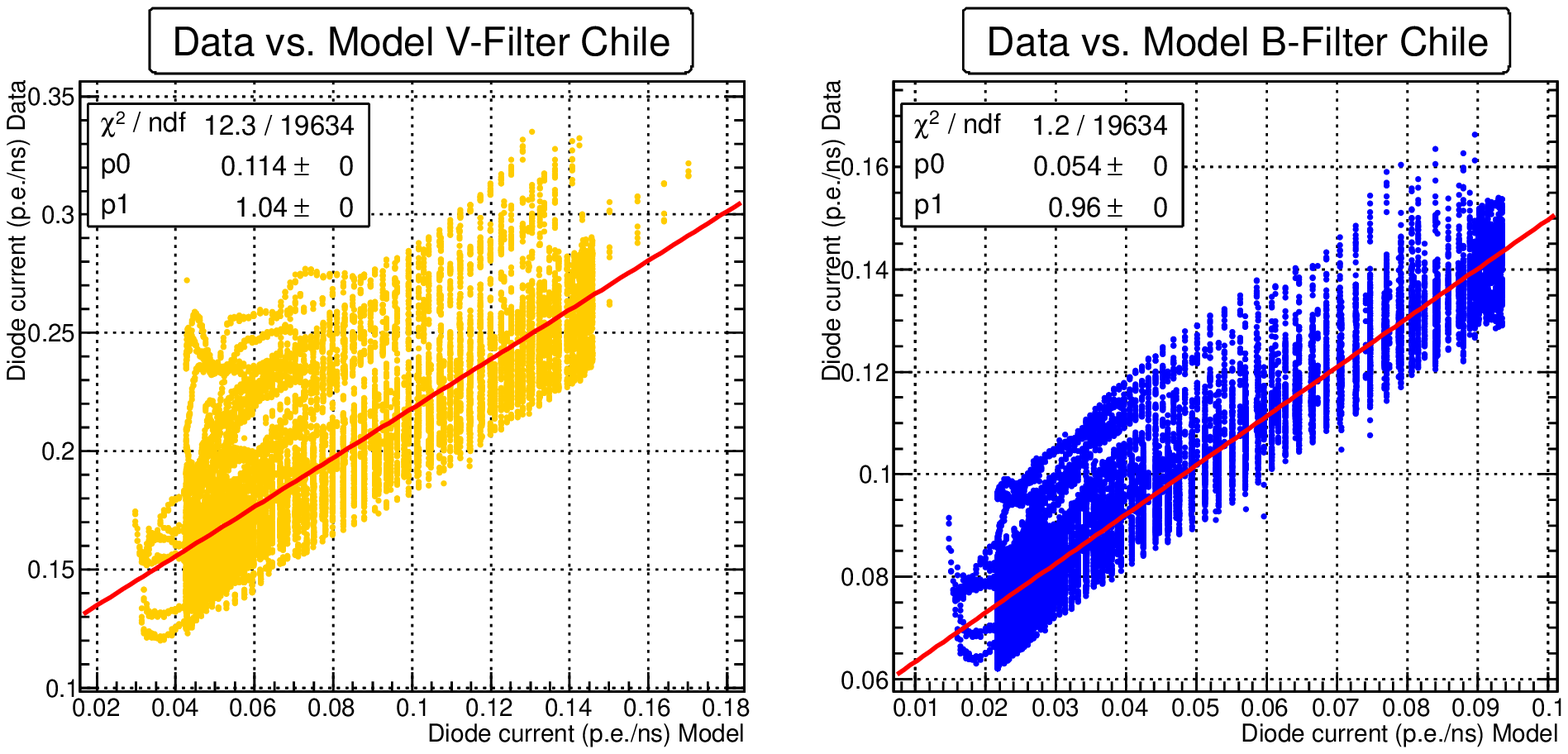}
\includegraphics[width=0.48\textwidth]{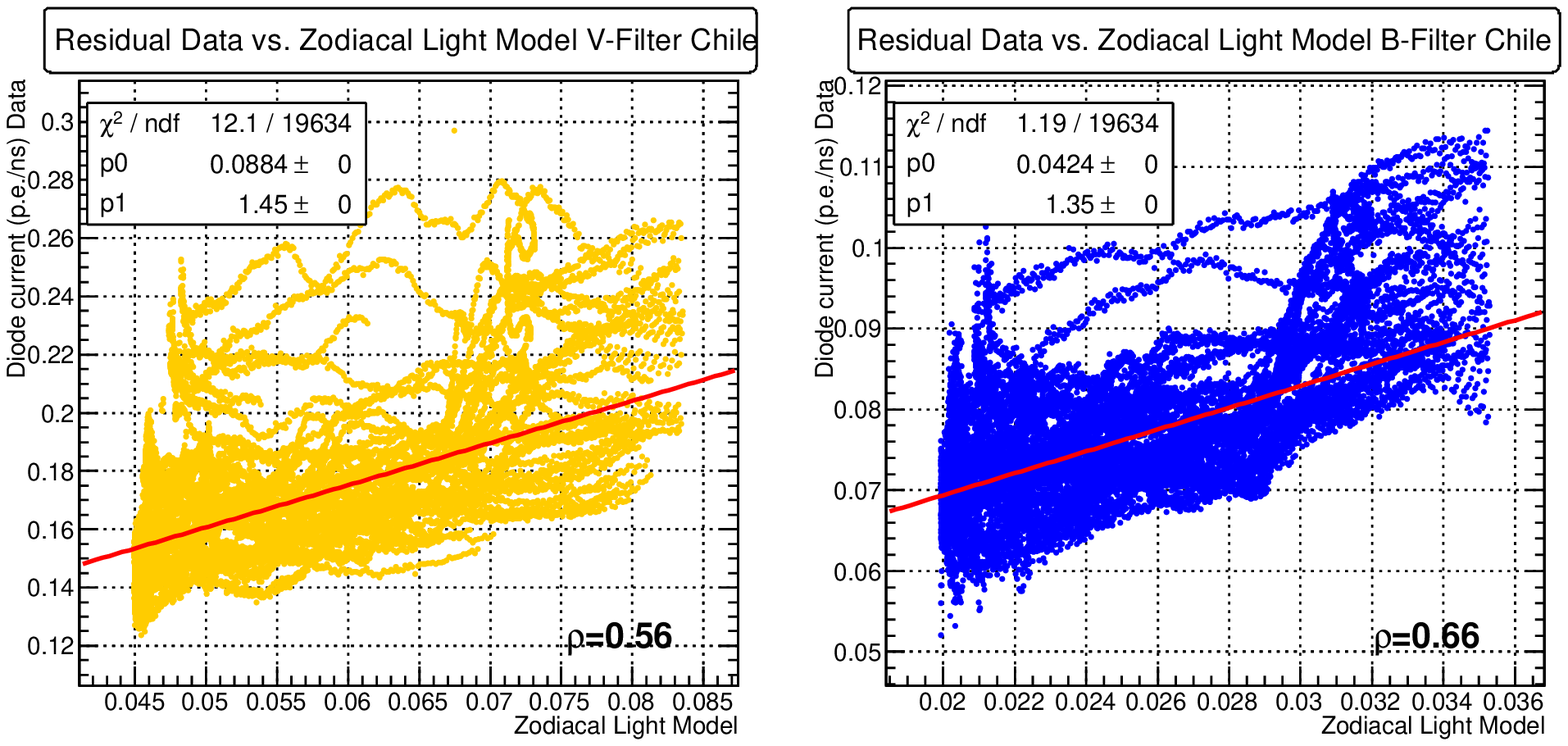}
\includegraphics[width=0.48\textwidth]{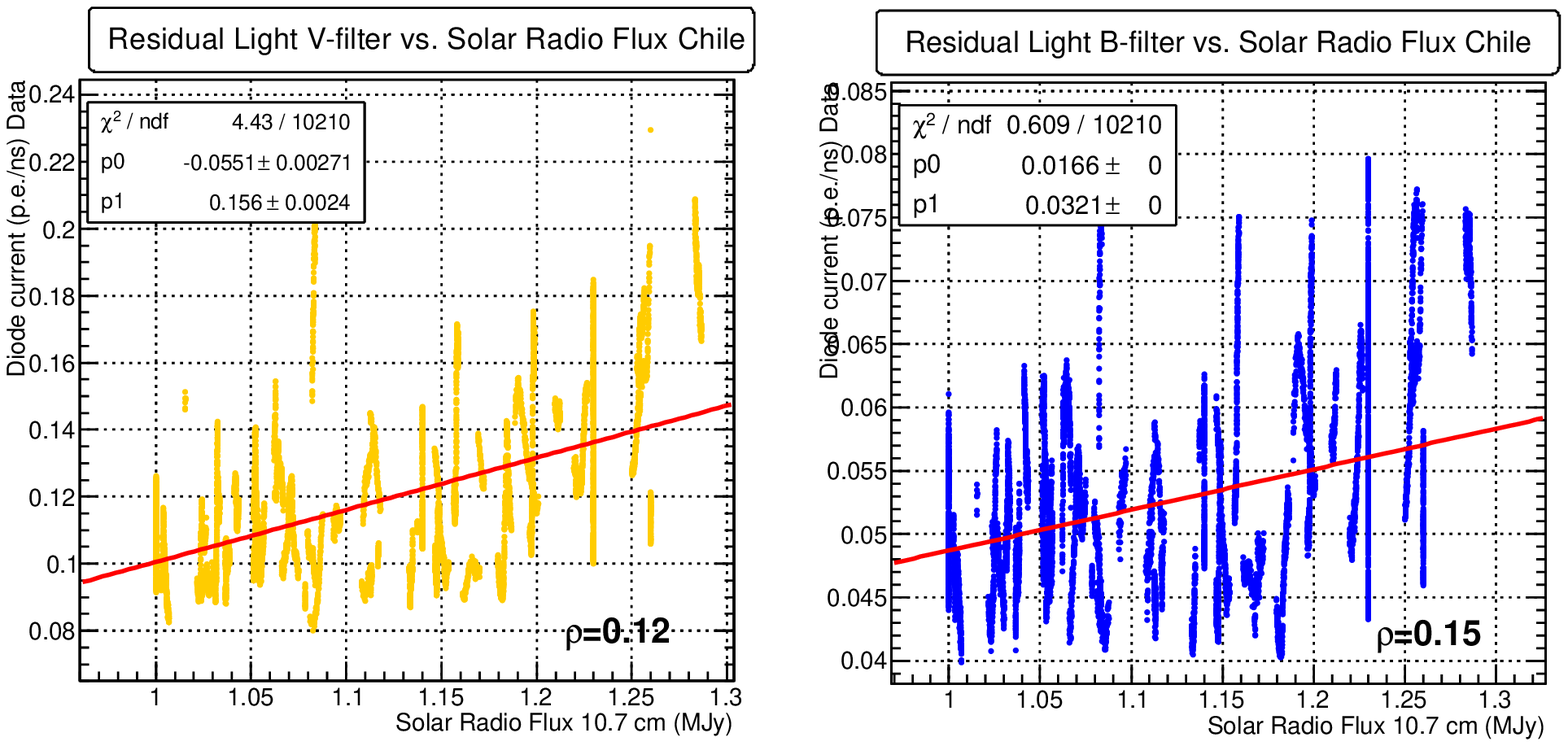}
\vspace{-0.8cm}
\caption{\small{Data from the atmoscope installed at Paranal, fitted against the starlight model (top), the zodiacal light model (after subtraction of starlight, center) and solar radio flux (bottom). \label{fits}
\vspace{-0.2cm}}}
\end{figure}

The contribution of zodiacal light to the light-of-night-sky can be significant. 
It has been modeled using the tables from~\cite{levasseur,leinert}, absorbed with modified atmospheric extinction coefficients 
from~\cite{noll}. The B-filter component was obtained assuming a solar spectrum and the correction of eq.~\ref{eq.bminb}, and 
the reddening prescription from~\cite{leinert}. Fig.~\ref{fits} (top) shows the obtained correlation between data and star light model, 
the central figure shows the correlation with the modeled zodiacal light component.

Airglow is emitted 
in the mesopause (at 90--100~km a.s.l.) and in the F2 layer of the 
ionosphere (at around 300~km a.s.l.). Its spectrum consists in the 300-750~nm range consists mainly of OH and O$_2$ molecular emission 
bands, permitted and forbidden lines of O, N and Na atom, emission lines originating from street lights and weak continuum emission~\cite{vladyuk}. 
The mesopause emits the quasi-continuum of the O$_2$ forbidden Herzberg bands, from 260--380~nm, 
the OI line at 557.7~nm and the NaD lines at 589.0 and 589.6~nm, respectively. 
The intensity of airglow correlates with the intensity of solar UV flux striking the upper atmosphere, which in turn correlates with the 
10.7~cm radio emission from the Sun and the number of sun spots~\cite{chatterjee}.

\section{Conclusions}

The atmoscopes are currently characterizing candidate sites for the CTA in the Northern and Southern hemisphere~\cite{bulik}, 
in a one-year site characterization campaign, with equal instrumentation, and a 
close follow-up of each individual instrument. This proceeding shows how the 
analysis of the light-of-night-sky part is made. Because of routinely made calibration updates, 
the statistical and systematic uncertainties of the measurements 
are estimated to be better than 15\%, splitting up each individual measurement point into its individual contributions 
can be made with the about same precision. 

\vspace{0.5cm}
{\bf Acknowledgements:} 
We gratefully acknowledge support from the agencies and organizations  
 listed in this page:\\
\url{http://www.cta-observatory.org/?q=node/22}

\bibliographystyle{plain}

\end{document}